\documentclass[conference]{IEEEtran}

\usepackage{cite}
\usepackage{amsmath,amssymb,amsfonts}
\usepackage{graphicx}
\usepackage{textcomp}
\usepackage{xcolor}
\usepackage{booktabs}
\usepackage{url}
\usepackage{tikz}
\usepackage{pgfplots}
\usepackage{pifont}
\usepackage{enumitem}
\usepackage[caption=false,font=footnotesize]{subfig}
\usepackage{todonotes}

\pgfplotsset{compat=1.17}
\usetikzlibrary{shapes.geometric,arrows.meta,positioning,fit,backgrounds,calc}

\newcommand{\cmark}{\ding{51}}
\newcommand{\xmark}{\ding{55}}
\newcommand{\namark}{--}

\begin{document}

\title{CERTIoT-6G: Continuous Cybersecurity Certification for IoT Devices in 5G/6G Networks}

\author{
  \IEEEauthorblockN{Evangelos Lempesis}
  \IEEEauthorblockA{Mulini}
  \and
  \IEEEauthorblockN{Fabio Palmese}  
  \IEEEauthorblockA{Mulini}
  \and
  \IEEEauthorblockN{Hamed Haddadi}
\IEEEauthorblockA{Mulini\\Imperial College London}
  \and
  \IEEEauthorblockN{Anna Maria Mandalari}
  \IEEEauthorblockA{Mulini\\University College London}

}

\maketitle

\begin{abstract}
The massive adoption of Internet of Things (IoT) devices across critical domains such as healthcare, smart cities, industrial automation, and critical infrastructure introduces significant cybersecurity and regulatory challenges. Current and forthcoming European regulations, including the Cyber Resilience Act (CRA) and the NIS2 Directive, require manufacturers, operators, and other organizations to ensure secure-by-design devices, continuous vulnerability management, and resilient operation throughout the device lifecycle. Traditional certification mechanisms remain static, manual, and difficult to scale across heterogeneous IoT ecosystems. This paper presents CERTIoT-6G, a Security-as-a-Service (SECaaS) framework that enables automated cybersecurity certification and continuous compliance monitoring of IoT devices operating in 5G and future 6G networks. The framework integrates automated compliance analysis, real-time traffic monitoring, and adversarial testing capabilities. We validate the CERTIoT-6G framework on different IoT device categories operating in an advanced 5G testbed. Evaluation results reveal critical compliance gaps, particularly in traffic encryption and availability under unstable conditions, and demonstrate that the framework produces actionable verdicts mapped to regulatory requirements across heterogeneous device types. Furthermore, we show that the monitoring pipeline has a negligible impact on live 5G traffic. %Finally, we outline how the framework architecture supports an incremental evolution toward AI-native compliance operations consistent with the broader 6G vision. 
\end{abstract}

% ═══════════════════════════════════════════════════════════════════════
\section{Introduction}
The exponential growth of the Internet of Things (IoT) is reshaping critical sectors including healthcare, smart cities, and industrial automation. Large numbers of heterogeneous connected devices generate continuous data streams that enable advanced monitoring and automation at scale. However, this same proliferation substantially expands the attack surface: vulnerable devices may expose sensitive personal information~\cite{ren2019}, allow unauthorized access, or serve as entry points for large-scale attacks. As IoT deployments increasingly rely on 5G~\cite{hasan2022} and future 6G infrastructures, the scale and dynamism of these ecosystems make manual security assessment both impractical and insufficient. European Union legislation has responded by establishing binding cybersecurity obligations for both manufacturers of connected products and organizations operating critical networked systems. The Cyber Resilience Act (CRA)~\cite{cra2025} mandates that manufacturers of connected products demonstrate secure-by-design principles, active vulnerability management, and compliance monitoring throughout the device lifecycle, not only at launch time. The NIS2 Directive extends these obligations to operators of essential services that depend on networked IoT deployments. Together, these frameworks create a regulatory environment in which certification must be continuous, evidence-based, and scalable to large device populations. 
Existing certification practices, however, remain largely manual and point-in-time. Standards such as ETSI\,EN\,303\,645~\cite{etsi303645} and its conformance specification ETSI TS 103~701 \cite{etsi_ts_103701} define the security requirements, yet the testing processes are typically performed once at product launch and do not track the evolving security posture of devices in operation. Recent work has shown that even with standard network security tooling, only 52--70\% of ETSI TS 103 701 tests can be automated~\cite{kaksonen2024}, highlighting the need for continuous monitoring solutions.\\
This paper introduces CERTIoT-6G, a framework that transforms IoT device certification from a static exercise into a continuous, automated service embedded within the 5G network infrastructure. Operating at the User Plane Function (UPF) via Local Breakout (LBO), the framework inspects live device traffic, executes adversarial compliance tests, and produces structured compliance reports mapped to specific CRA and NIS2 requirements, without requiring device software manipulation. We validate the framework on commercial IoT devices operating on a 5G network, uncovering significant compliance gaps and demonstrating that automated, network-embedded certification can produce compliance evidence at scale. The main contributions of this paper are:
\begin{itemize}[leftmargin=3mm]
  \item A Security-as-a-Service (SECaaS) architecture enabling continuous cybersecurity certification of IoT devices in 5G and future 6G networks.
  \item An automated compliance monitoring pipeline translating regulatory frameworks (CRA, NIS2, ETSI EN 303~645) into network-level compliance tests with per-requirement verdicts.
  \item Integration within an advanced 5G Standalone (SA) testbed, demonstrating real-time traffic monitoring via UPF/LBO.
  \item An early experimental validation across three representative device categories revealing critical gaps in traffic encryption and denial-of-service resilience, with planned extension to a larger and more diverse device population.
\end{itemize}

% ═══════════════════════════════════════════════════════════════════════
\section{Related Work}

\subsection{IoT Security Measurement and Evaluation}

Early systematic efforts to evaluate the security of IoT deployments revealed widespread and structural weaknesses. Alrawi et al.~\cite{alrawi2019} systematized the attack surface of home-based IoT across firmware, network, application, and cloud components, showing that vulnerabilities span the entire deployment stack. Ren et al.~\cite{ren2019} quantified the breadth of personal data exposed through unencrypted or insufficiently protected traffic in consumer IoT devices, findings strongly corroborated by our experimental results. Huang et al.~\cite{huang2020} demonstrated that gateway-level traffic collection can yield labeled datasets suitable for large-scale security and privacy analysis of smart-home devices, and Shahid et al.~\cite{shahid2018} showed that passive traffic features alone are sufficient to identify IoT devices by type and vendor without device-side access.

\subsection{Compliance-Oriented IoT Security Frameworks}

Most directly related to CERTIoT-6G is COPSEC~\cite{copsec2023}, a compliance-oriented framework that automatically probes IoT devices at the network gateway and evaluates them against security guidelines and privacy regulations. COPSEC extracts measurable metrics from standards including ETSI EN 303~645 and ENISA guidelines, applies them through automated network experiments, and reports compliance verdicts. CERTIoT-6G shares COPSEC's gateway-level measurement paradigm and regulatory grounding, but extends it in three critical directions: it operates within a live 5G network infrastructure, leveraging the UPF and LBO rather than a local Wi-Fi gateway, it incorporates active adversarial testing beyond passive traffic analysis, and it maps each verdict directly to individual CRA and NIS2 requirements for regulatory auditability.

\subsection{Automated Security Testing Standards}

ETSI TS 103~701~\cite{etsi_ts_103701} defines a conformance assessment methodology for consumer IoT devices against ETSI EN 303~645. While comprehensive, its procedure is designed for manual execution by accredited testing laboratories. Kaksonen et al.~\cite{kaksonen2024} investigated the degree to which standard network security tools can automate the 56 test units relevant to network attacks within ETSI TS 103~701, finding 52\% coverage with basic tooling and up to 70\% with advanced tools, leaving a significant automation gap. CERTIoT-6G addresses this by embedding a curated set of automated compliance tests, covering service exposure, encryption, authentication, data leakage, and resilience directly within the 5G network plane, enabling continuous coverage aligned with CRA requirements.

\subsection{IoT Security in 5G and 6G Networks}

The adoption of cellular connectivity in IoT deployments introduces specific security considerations. Hasan et al.~\cite{hasan2022} provide a comprehensive survey of threats for 5G-enabled Internet of Medical Things, identifying credential attacks, protocol vulnerabilities, and denial-of-service as primary threats, all of which are exercised by the CERTIoT-6G validation engine. The 3rd Generation Partnership Project (3GPP) system architecture for 5G~\cite{3gpp2020} provides the UPF and LBO mechanisms that CERTIoT-6G exploits to position its monitoring functions at the network edge. As 6G standardisation advances, AI-native network management is expected to become a core design principle, creating further demand for adaptive compliance services beyond what static certification can provide.

\subsection{Positioning of CERTIoT-6G}

Unlike prior measurement tools that operate on local networks and evaluate devices passively~\cite{copsec2023,huang2020}, and unlike testing frameworks designed for laboratory execution against a conformance specification~\cite{kaksonen2024,etsi_ts_103701}, CERTIoT-6G operates continuously inside a live 5G network infrastructure. It combines passive monitoring, active adversarial testing, and natural language processing (NLP)-based personal-data detection into a single pipeline that produces machine-readable compliance evidence bound to specific regulatory requirements, a capability not provided by existing tools.

% ═══════════════════════════════════════════════════════════════════════
\section{CERTIoT-6G Framework}

\begin{figure}[t]
\centering
\includegraphics[width=\linewidth]{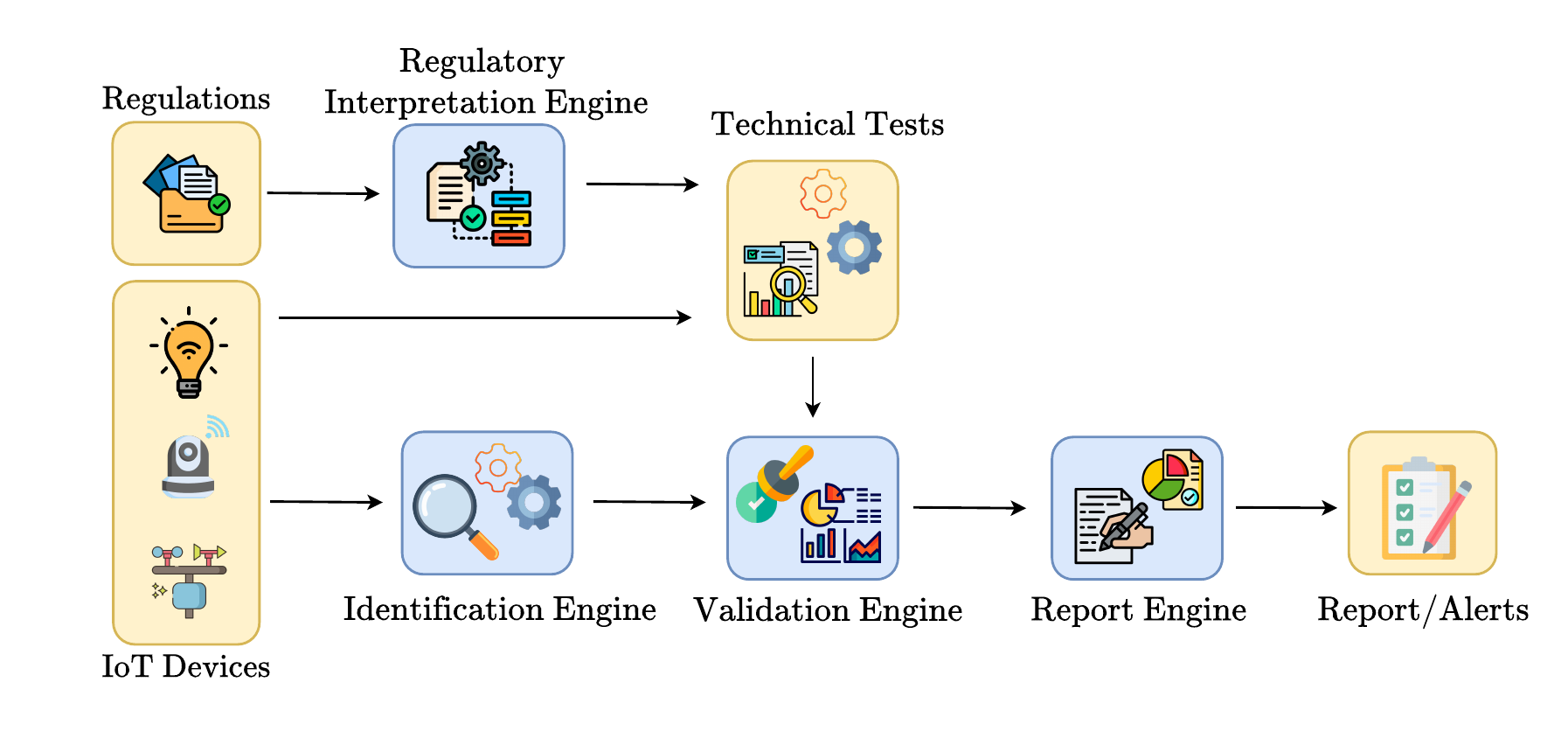}
\caption{CERTIoT-6G four-engine architecture. The Regulatory Interpretation Engine maps standards to test bindings; the Device Identification Engine auto-registers devices; the Validation Engine runs passive and active
tests; the Reporting Engine aggregates verdicts into compliance reports.}
\label{fig:architecture}
\end{figure}

CERTIoT-6G is a SECaaS platform that continuously monitors IoT devices connected to 5G or future 6G networks and verifies their compliance with applicable cybersecurity regulations. Figure~\ref{fig:architecture} illustrates the four functional components and the data flows between them.

% ── Figure 1: Architecture diagram ──────────────────────────────────
% ─────────────────────────────────────────────────────────────────────

\subsection{Regulatory Interpretation Engine}
The interpretation engine decomposes cybersecurity regulations into structured, machine-readable requirement bindings. Frameworks and guidance - including the CRA~\cite{cra2025}, the NIS2 Directive \cite{nis2_directive_2022}, ETSI EN 303~645~\cite{etsi303645}, together with ENISA guidelines~\cite{enisa2020} and NIST recommendations~\cite{nist2024} - are mapped to concrete security tests so that each result is bound to one or more specific requirements. The mapping is many-to-many: a single test can provide evidence for multiple requirements, allowing the framework to amortise testing effort across overlapping regulatory obligations.

\subsection{Device Identification Engine}
%The identification engine discovers and registers devices as they join the network, relying on DHCP hooks and passive traffic sniffing to extract the MAC address, hardware vendor, and hostname~\cite{shahid2018}. Each time a new device is detected in the network, it is immediately handled by the identification engine to obtain the required information: category, vendor, model, and version
The identification engine discovers and registers devices as they join the network, relying on passive traffic sniffing. Each time a new device is detected on the network, the identification module is triggered to obtain the required device information: category, vendor, model, and version. In our case, the identification pipeline is based on the method described in \cite{ciechonski2026early}. The first 30 seconds of device traffic after its detection is extracted, and DNS-based statistical features serve as input to a machine learning classifier that determines the device type. The derived device attributes allow the system to associate each device with the appropriate set of compliance checks and enable device-class-specific policies without manual onboarding.

\subsection{Validation Engine}
The validation engine instantiates and executes the test suite produced by the Regulatory Interpretation Engine, receiving a structured set of test bindings that associate each compliance check with one or more specific regulatory requirements.  The test suite is not automatically generated: each test is defined and validated by security experts with knowledge of both the relevant regulatory requirements and the network-level mechanisms through which those requirements can be violated, ensuring that verdicts are legally defensible and technically grounded. Table \ref{tab:tests} summarizes part of the test suite, derived from a thorough analysis of recent regulations. The table also reports the binding of the tests to the ETSI EN 303 645 standard, from which many of the CRA standards are derived. Tests are executed in two complementary modes: \textit{passive traffic monitoring}, which observes device behavior without interfering with it, and \textit{active security probing}, which subjects the device to controlled adversarial activity and measures its response.

\textbf{Passive monitoring.}
The device traffic is captured using standard tools (e.g., tshark or tcpdump) and classified by application-layer protocol to compute the proportion of encrypted versus cleartext flows. This drives two tests:

\begin{itemize}[leftmargin=5mm]
  \item \textit{Encrypted Traffic}: measures what fraction of device flows use a recognised encryption protocol. Devices transmitting part of their data in cleartext are flagged as non-compliant with confidentiality requirements.
  \item \textit{Personal Information}: captured payloads are scanned by a pre-trained NLP entity-recognition model to detect personally identifiable information (PII) such as names, locations, device identifiers, and similar attributes transmitted without adequate protection.
\end{itemize}
These tests are entirely passive and do not interfere with normal operation of the device. The duration of each test is inherently device-dependent: an adequate observation window is essential for accurate verdicts, since a window that is too short may fail to capture the full range of a device's communication behavior and produce misleading results.

\textbf{Active probing.}
The engine executes a sequence of targeted adversarial tests:

\begin{itemize}[leftmargin=5mm]
  \item \textit{Unused Ports}: a TCP/UDP port scan enumerates open services and compares them against the services observed in passive captures; ports that are open but absent from legitimate traffic are flagged as unnecessary attack surface.
  \item \textit{TLS Interception}: an intercepting proxy is inserted between the device and its cloud endpoint to assess whether the device validates the server certificate chain and rejects connections through an untrusted intermediary. This verifies the device resilience to Man-in-the-Middle attacks (MITM).
  \item \textit{Dictionary Attack}: a credential brute-forcing sequence is run against exposed authentication interfaces (e.g., SSH, HTTP); the test checks whether the device enforces rate limiting or account lockout under repeated failed attempts.
  \item \textit{Replay Attack}: previously captured valid traffic frames are replayed to assess whether the device implements message-freshness mechanisms (e.g., nonces, sequence numbers) sufficient to reject replayed commands.
  \item \textit{DoS Attack} and \textit{Reliability}: a flood attack is directed at the device for a controlled duration; the test records whether the device remains responsive during the attack and measures the time to full service recovery afterwards, combining availability and resilience into a paired verdict.
\end{itemize}

Each test produces a binary verdict as compliant or non-compliant, together with the triggering evidence, and passes the result to the Reporting Engine.%with the regulatory requirement binding inherited from the Interpretation Engine.

\begin{table}[t]
\caption{CERTIoT-6G Tests with Targeted Properties and Requirements}
\label{tab:tests}
\centering\small
\begin{tabular}{@{}llll@{}}
\toprule
\textbf{Test} & \textbf{Property} & \textbf{EN 303 645 Req.}\\
\midrule
Unused Ports (Port) & Attack Surface & 5.6.1   \\
TLS Interception (TLS) & Confidentiality & 5.1. 5.3, 5.5\\
Encrypted Traffic (Enc) & Confidentiality & 5.1, 5.3, 5.5, 5.8\\
Dictionary Attack (Dict) & Authentication & 5.1   \\
Replay Attack (Rep) & Integrity & 5.5       \\
DoS Attack (DoS)        & Availability & 5.9.3     \\
Personal Info (PII)  & Privacy & 5.8 \\
Reliability (Rel) & Resilience  & 5.9.3     \\
\bottomrule
\end{tabular}
\end{table}

\subsection{Reporting Engine}
The reporting engine generates structured compliance reports and user-friendly web-based alerts when non-compliance is detected. Rather than enforcing network-level mitigations - which may not be feasible for every topology or installation - the engine documents the non-compliance, classifies its severity, and provides actionable remediation recommendations such as closing insecure ports, restricting exposed services, or applying firmware updates, keeping remediation authority with device owners and operators. Each test yields a binary compliant or non-compliant verdict with a human-readable rationale, a proposed remediation step, and the bound regulatory requirement; results are aggregated into reports that can support certification, audit, and conformity-assessment processes.

% ═══════════════════════════════════════════════════════════════════════
\section{Integration in 5G Testbeds}
\begin{figure}[t]
\centering
\includegraphics[width=0.96\linewidth]{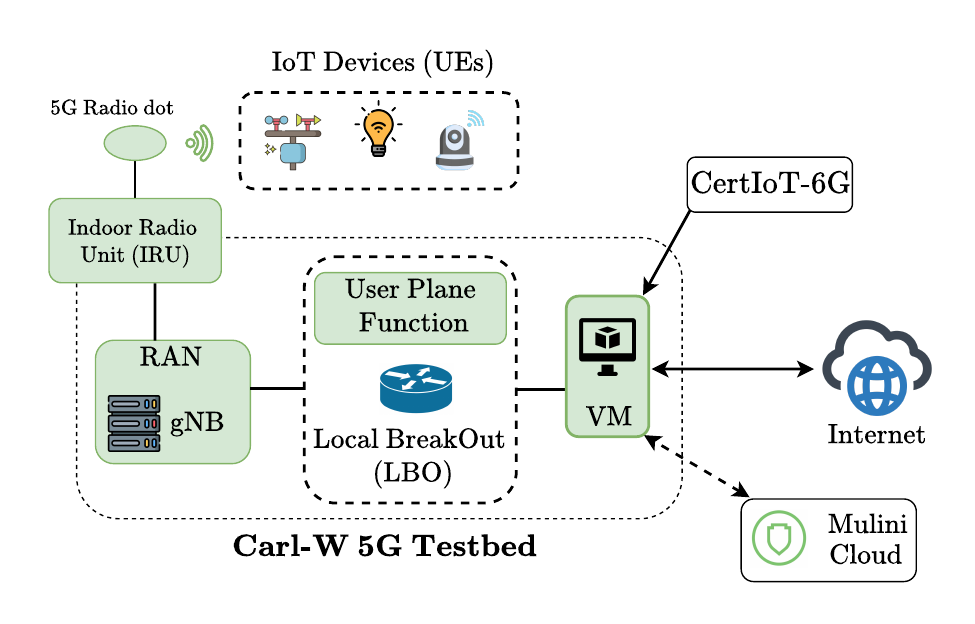}
\caption{CERTIoT-6G deployment in the KAU CARL-W 5G testbed. IoT devices connect directly to the 5G network as UEs. Traffic is inspected at the analysis VM via LBO; the Mulini cloud is reached over WireGuard VPN.}
\label{fig:testbed}
\end{figure}

To validate the proposed solution in a real-world operational 5G testbed, we deploy CERTIoT-6G in the CARL-W testbed at Karlstad University (KAU)\footnote{https://6gpath.eu/testbeds/karlstad-university-testbed/}, leveraging its 5G SA infrastructure for network-level IoT evaluation~\cite{carl-w}.\\
The key integration point is the UPF with LBO ~\cite{3gpp2020}. In this configuration, the user-plane traffic generated by the IoT devices is locally steered toward a dedicated analysis virtual machine (VM) before being forwarded to the Internet. This allows the monitoring platform to inspect device communications in-path, close to the network edge, while avoiding unnecessary traversal of remote cloud infrastructure and minimizing additional latency.\\
%
%The experimental setup follows the split architecture shown in Figure~\ref{fig:testbed}. IoT devices connect directly to the 5G network as User Equipment (UE). Their traffic is then routed through the 5G user plane and, via LBO, delivered to an edge VM hosting the CERTIoT-6G monitoring components. On this VM, a remote capture agent based on \texttt{tshark} records device traffic on demand, while the compliance analysis modules perform protocol classification and personal-data inspection. During the inspection, the traffic is forwarded toward the Internet, preserving normal device connectivity while enabling continuous network-level compliance monitoring.
%
The deployment architecture is shown in Figure~\ref{fig:testbed}. IoT devices attach to the CARL-W 5G SA network as UEs \cite{kau-testbed-analysis}. On the radio side, the testbed relies on the CARL-W Radio Access Network infrastructure, including Ericsson radio dots, indoor radio units, and the gNodeB. User-plane traffic is then carried through the 5G data path and anchored at the locally deployed UPF in Karlstad. Through LBO, selected traffic is steered to the CERTIoT-6G VM deployed in the KAU edge environment before being forwarded to the Internet.
This placement allows CERTIoT-6G to inspect live IoT traffic, close to the 5G edge, without requiring modifications to the devices. The VM hosts the remote capture agent and the compliance analysis modules, including protocol classification, personal-data inspection, and the active security tests.
Secure connectivity between the Mulini cloud infrastructure and the KAU testbed is maintained through WireGuard VPN tunnels. This enables coordination between the cloud-side management components and the edge-deployed monitoring VM, supporting synchronised monitoring and reporting across both environments.\\

% ── Figure 2: Testbed deployment diagram ────────────────────────────
% ─────────────────────────────────────────────────────────────────────

% ═══════════════════════════════════════════════════════════════════════
\section{Experimental Validation}

\subsection{Setup and Device Set}
As an early validation of the framework, three representative IoT devices were selected to cover a range of communication patterns, data sensitivity levels, and service exposure characteristics, belonging to three distinct categories:\textit{surveillance}, \textit{lighting}, and \textit{environment}. These categories reflect typical IoT deployments in smart-home and smart-city scenarios and present meaningfully different compliance profiles. The devices operated as 5G User Equipment (UE) devices, with their traffic forwarded to the analysis VM via the 5G network and LBO path. Each device was subjected to the full battery of eight compliance tests listed in Table~\ref{tab:tests}, where applicable. The Replay Attack test was executed only for devices that expose locally reachable services; the Dictionary Attack test was applied only to devices that present authentication interfaces. Extension of the validation to a larger and more diverse device population is planned for future work.

\subsection{Analysis and Key Findings}

Table~\ref{tab:results} reports the compliance verdict for each device category across all applicable tests (\cmark\ means compliant, \xmark\ non-compliant, and \namark\ not applicable).
\begin{table}[t]
\caption{Compliance verdicts across the three device categories.}
\label{tab:results}
\centering
\setlength{\tabcolsep}{3pt}
\scriptsize
\begin{tabular}{@{}lcccccccc@{}}
\toprule
\textbf{Device} & \textbf{TLS} & \textbf{Enc} & \textbf{Port}
  & \textbf{PII} & \textbf{Rel} & \textbf{Dict} & \textbf{DoS} & \textbf{Rep}\\
\midrule
Surveillance    &\cmark&\xmark&\cmark&\xmark&\cmark&\namark&\xmark&\cmark\\
Lighting         &\xmark&\xmark&\cmark&\cmark&\cmark&\cmark &\xmark&\xmark\\
Environment       &\cmark&\xmark&\cmark&\cmark&\cmark&\namark&\cmark&\cmark\\
\midrule
\textbf{Pass rate} & 66.7\% & 0\% & 100\% & 66.7\% & 100\% & 100\% & 33.3\% & 66.7\%\\
\bottomrule
\end{tabular}
\end{table}
The last row of the table summarises per-test compliance rates across the three device categories. Despite the small initial device set, the results reveal a clear and consistent disparity between security properties, establishing a baseline picture that motivates the planned broader evaluation.

\textbf{Strengths.}
All three devices (100\%) recovered to full operation after the DoS probing phase (\textit{Reliability}), and the single device presenting an authentication interface (the smart light) resisted credential brute-forcing (\textit{Dictionary Attack} 100\%). Port hygiene was consistent across all categories: no device exposed unnecessary network services (\textit{Unused Ports} 100\%), confirming that basic attack surface reduction is implemented even at the lower end of the device spectrum.

\textbf{Critical failure: traffic encryption.}
\textit{Encrypted Traffic} was the most critical compliance failure as all the tested devices transmitted at least part of their information in cleartext, exposing data to passive eavesdropping. This finding corroborates patterns documented in prior network measurement studies~\cite{ren2019,copsec2023} and constitutes a direct violation of many cybersecurity requirements for confidentiality of data in transit. The fact that this failure appears in both a media-rich device (camera) and a lightweight constrained device (lighting) suggests it is not specific to a single device class.

\textbf{Availability and integrity.}
Only the environmental sensor passed the \textit{DoS Attack} test, indicating that the camera and lighting device fail to maintain service continuity under flooding conditions, a concern for deployments where availability carries safety implications. For the \textit{Replay Attack}, the camera and environmental sensor passed, as their services rejected replayed messages, while the lighting device failed, allowing unauthorized entities to replicate previously sniffed on/off commands.

\textbf{Privacy.} Two of three devices (67\%) passed the \textit{Personal Information} test. The surveillance camera transmitted personal data in its network payloads, a foreseeable but potentially non-compliant finding under the mapped confidentiality requirements. Two of three devices (67\%) also passed the \textit{TLS Interception} test: the lighting device accepted traffic relayed through an intercepting proxy, indicating an absent or ineffective certificate validation, while both the surveillance camera and the environmental sensor resisted the man-in-the-middle interception.

\textbf{Device-level summary.}
The environmental sensor showed the strongest compliance profile, passing six of the seven applicable tests. The surveillance camera showed a mixed profile: strong on TLS and replay integrity, but failing on encryption and privacy, findings directly linked to the nature of the data it handles. The smart light showed the weakest overall compliance, failing TLS interception, encryption, DoS, and replay, which is consistent with the lower security investment typically found in constrained illumination devices. These preliminary results, while limited in sample size, already reveal compliance patterns that are device-class-specific and regulatorily actionable; extending the evaluation to a larger and more diverse device set will determine how representative these patterns are.

% ═══════════════════════════════════════════════════════════════════════
% ── Figure 4
\begin{figure}[t]
\centering

\includegraphics[width=0.94\linewidth]{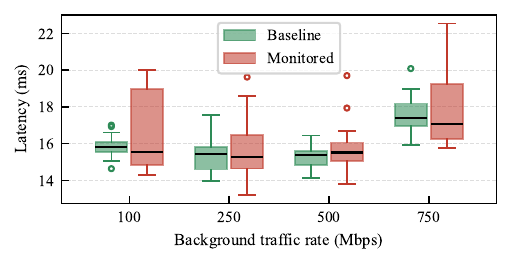}

%\vspace{0.5em}

\includegraphics[width=0.94\linewidth]{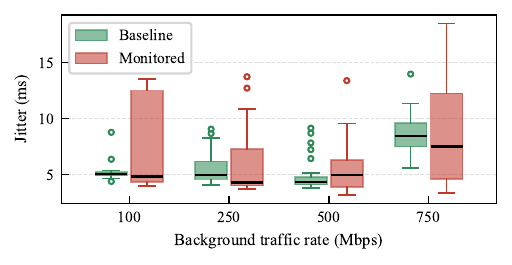}

\caption{Box plot of the Mean RTT (latency, top) and RTT standard deviation (jitter,
bottom) under baseline and monitored conditions across 100 runs.}
\label{fig:network_impact}
\end{figure}

\section{Network Impact and Future Directions}

\subsection{Network Impact Evaluation}

To quantify the overhead introduced by the CERTIoT-6G monitoring pipeline on the live 5G network, we designed a controlled two-phase experiment in the KAU CARL-W testbed. The goal is to determine whether running the full security test suite alongside production traffic measurably degrades key network quality indicators compared to a clean baseline.
Background traffic is generated using \textit{iperf3}\cite{iperf3}, an open-source network performance tool that operates in a client-server model and supports multi-stream TCP/UDP flows with configurable bandwidth and duration. Its ability to produce controlled, reproducible synthetic loads makes it well suited for isolating the impact of a monitoring tool on ambient network conditions.

\textbf{Phase~1 --- Baseline.}
An iperf3 client generates background traffic toward the testbed server running on the VM using five parallel TCP streams with varying transmission rates (100 Mbps to 750 Mbps total rate) for 60 seconds. Simultaneously, tcpdump captures traffic on the network interface to collect all the iperf3-generated packets, filtering on port 5201 with a snapshot length of 96 bytes to collect header-only traces. During this phase the CERTIoT-6G monitoring tool is completely idle.

\textbf{Phase~2 --- Monitored.}
The same iperf3 session runs continuously in the background while the full CERTIoT-6G test suite executes sequentially in the foreground, in the following order: unused ports scan, encrypted traffic analysis, personal information inspection, replay attack, dictionary attack, recovery test, DoS attack, and TLS interception. A single tcpdump capture covers the entire monitored period. In both phases, the captured traces are analyzed offline using \textit{tstat}, which extracts TCP-level statistics per flow, coupled with the iperf3 server output parameters.

Four network quality indicators are compared between the two phases: mean round-trip time (RTT) as a measure of latency impact, standard deviation of the RTT as a measure of delay variation (jitter), packet retransmission rate, and throughput. The full two-phase experiment is repeated 100 times; we report mean values and standard deviation across runs to characterize both central tendency and variability.

\begin{figure}[t]
\centering
\includegraphics[width=0.94\linewidth]{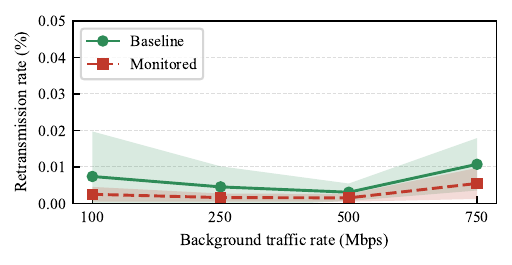}
\caption{Packet retransmission rate under baseline and monitored conditions.}
\label{fig:retransm}
\end{figure}

Figures~\ref{fig:network_impact} and \ref{fig:retransm} present the results.
The monitoring framework introduces minimal overhead on the 5G network. Mean RTT and jitter both remain within acceptable bounds, with a maximum observed increase below 10 ms between baseline and monitored conditions. No application-level failures were observed, and TCP retransmission rates remained very low in both phases, with no meaningful difference between baseline and monitored conditions. Measured throughput exactly matched the configured iperf3 transmission rate in all runs, reason for which we omit to report the values. These results confirm that CERTIoT-6G can operate continuously alongside production 5G traffic without materially degrading network quality.

% ─────────────────────────────────────────────────────────────────────

\subsection{Benefits to the Certification Lifecycle}
An analysis of existing platforms for device compliance certification reveals that the process is traditionally performed manually and can require several months for full device compliance certification. Automated compliance assessment enables scaling to larger device populations, reduces per-device cost, and compresses compliance decision time from months to under 48 hours, a crucial capability under the CRA, which requires manufacturers to respond to newly discovered vulnerabilities within defined timescales. For example, part of the test suite provided in this work can run and produce the compliance verdict in less than two minutes, highlighting the efficiency of the proposed solution for large-scale implementations. Furthermore, the capability to produce a decision with limited supervision reduces human effort and thus the verification costs.

%\begin{table}[t]
%\caption{Expected Benefits over Conventional Certification}
%\label{tab:benefits}
%\centering\small
%\begin{tabular}{@{}lcc@{}}
%\toprule
%\textbf{Aspect} & \textbf{Baseline} & \textbf{Target} \\
%\midrule
%Device Coverage       & 2       & $\geq$20          \\
%Certification Cost    & 100\%   & $\leq$70\%        \\
%Time to Compliance    & Months  & $<$48\,hours      \\
%\bottomrule
%\end{tabular}
%\end{table}

%\subsection{Stakeholder Ecosystem}
%CERTIoT-6G is validated in collaboration with four external
%stakeholders: Unione Montana Comuni Olimpici Via Lattea (smart-city
%pilot), Vodafone IT (industrial and commercialisation feedback), the
%Italian National Cybersecurity Agency ACN (CRA alignment), and the
%Italian Data Protection Authority GPDP (privacy implications of the
%monitoring functions).

\subsection{Toward AI-Native Compliance Operations}
The current validation engine relies on deterministic rule-based traffic analysis complemented by a pre-trained NLP model for personal-data detection. As the framework evolves toward future 6G deployments, several components are natural candidates for learning-based extension: (i) anomaly-driven detection to complement fixed classification thresholds under highly dynamic traffic conditions, and (ii) automated mapping of new or amended regulatory text to structured test bindings. We envision this evolution as incremental, preserving the auditable rule-based core while selectively extending it where learning-based methods offer measurable benefit.

% ═══════════════════════════════════════════════════════════════════════
\section{Conclusion}

This paper presented CERTIoT-6G, a Security-as-a-Service framework for automated, continuous cybersecurity certification of IoT devices in 5G and future 6G networks. By integrating regulatory interpretation, passive traffic monitoring, active adversarial testing, and structured compliance reporting into a unified pipeline deployed at the 5G UPF, the framework addresses the scalability and continuity gaps that conventional certification cannot bridge in dynamic IoT ecosystems.

An early validation across three representative device categories (surveillance, lighting, environment) in the CARL-W 5G SA testbed demonstrated end-to-end compliance monitoring across eight security test categories aligned with the CRA and the NIS2 Directive. Results revealed critical compliance gaps in traffic encryption and DoS resilience, alongside consistent strength in device post-attack recovery, port hygiene, and authentication security. A controlled two-phase network impact evaluation repeated 100 times measured the framework's overhead on mean RTT, jitter, packet retransmissions, and throughput. No application-layer loss was observed in either phase, and the maximum latency increase between baseline and monitored conditions remained below 10\,ms, confirming that continuous certification can operate alongside production 5G traffic without material degradation of network quality.
Looking forward, the validation will be extended to a larger and more diverse device population to establish statistically robust compliance baselines across device classes and manufacturers. The framework will incorporate AI-native components, adaptive testing, and large language model (LLM)-assisted regulatory mapping, consistent with the broader vision of 6G as a natively intelligent infrastructure.
\section*{Acknowledgements}
This work was partially supported by the EU Horizon Europe Program under Grant No. 101293242 (PIONEERS-6G) and the EU Horizon Europe SNS JU project under Grant No. 101139172 (6G-PATH).
% ═══════════════════════════════════════════════════════════════════════
\vspace{-2mm}
\bibliographystyle{IEEEtran}
\bibliography{references}
\vspace{-2mm}
\end{document}